\documentclass[sigconf,nonacm ]{acmart}
\setcopyright{cc}
\setcctype[4.0]{by}
\acmConference[IoT '25]{15th International Conference on the Internet of Things}{Nov 19--21,
  2025}{Vienna, AT}
\acmISBN{978-1-4503-XXXX-X/2018/06}

\usepackage{graphicx}
\usepackage{placeins}
\usepackage{hyperref}
\usepackage{slashbox}
\usepackage{amsmath}
\usepackage{pifont}
\usepackage{float}
\usepackage{llncsconf}
\usepackage{array}

\usepackage{multirow}

\usepackage{multicol}
\usepackage[per-mode=symbol]{siunitx}
\usepackage{comment}
\usepackage{siunitx}

\usepackage{siunitx}
\usepackage{url}

\usepackage{amsmath,amsfonts} 
\usepackage{algorithmic}
\usepackage{textcomp}
\usepackage{xcolor}

\usepackage{graphicx}
\usepackage{subcaption}
\usepackage{cleveref}
\usepackage{lipsum}

\def\BibTeX{{\rm B\kern-.05em{\sc i\kern-.025em b}\kern-.08em
    T\kern-.1667em\lower.7ex\hbox{E}\kern-.125emX}}

\begin{document}

\title{Wi-Fi Enabled Edge Intelligence Framework for Smart City Traffic Monitoring using Low-Power IoT Cameras}



\makeatother


\author{Raphael Walcher}
\email{{rawalcher}@edu.aau.at}
\affiliation{\institution{University of Klagenfurt}
             \department{Institute of Information Technology}
             \city{Klagenfurt}
             \country{Austria}}

\author{Kurt Horvath}
\email{{kurt}.{horvath}@aau.at}
\affiliation{\institution{University of Klagenfurt}
             \department{Institute of Information Technology}
             \city{Klagenfurt}
             \country{Austria}}

\author{Dragi Kimovski}
\email{{dragi}.{kimovski}@aau.at}
\affiliation{\institution{University of Klagenfurt}
             \department{Institute of Information Technology}
             \city{Klagenfurt}
             \country{Austria}}

\author{Stojan Kitanov}
\email{{stojan}.{kitanov}@unt.edu.mk}
\affiliation{\institution{Mother Teresa University}
             \department{Faculty of Information Sciences}
             \city{Skopje}
             \country{North Macedonia}}

\begin{abstract}
Real-time traffic monitoring in smart cities demands ultra-low latency processing to support time-critical decisions such as incident detection and congestion management. While cloud-based solutions offer robust computation, their inherent latency limits their applicability for such tasks. This work proposes a localized edge AI framework that connects low-power IoT camera sensors to a client, or applies offloading of inference to an NVIDIA Jetson Nano (GPU). Networking is achieved via Wi-Fi, enabling image classification without relying on wide-area infrastructure such as 5G, or wired networks. We evaluate two processing strategies: local inference on camera nodes and GPU-accelerated offloading to the Jetson Nano. We show that local processing is only feasible for lightweight models and low frame rates, whereas offloading enables near-real-time performance even for more complex models. These results demonstrate the viability of cost-effective, Wi-Fi-based edge AI deployments for latency-critical urban monitoring.

\end{abstract}



 \maketitle

\setlength{\abovedisplayskip}{4pt}
\setlength{\belowdisplayskip}{4pt}
\setlength{\abovedisplayshortskip}{4pt}
\setlength{\belowdisplayshortskip}{4pt}

\section{Introduction}\label{sec:Introduction}


Smart cities function as intelligent systems designed to enhance quality of life by addressing key areas such as mobility and traffic monitoring \cite{alahi2023integration}, as well as energy optimization and waste management. Among these applications of smart cities, real-time traffic analysis is central in enabling timely decision-making \cite{barthelemy2019edge} \cite{ning2019vehicular} for incident detection, congestion mitigation, and adaptive signal control. Such applications require ultra-low latency and a large amount of compute resources to be effective \cite{horvath20256g}. While traditional cloud-based solutions ensure substantial computational power, their inherent network latency and limited Quality of Service (QoS) guarantees pose significant challenges for supporting latency-critical and real-time smart city applications.

Emerging paradigms in edge computing and artificial intelligence (AI) address this challenge by bringing computation closer to the data source \cite{Kimovski21}, thereby reducing latency and bandwidth consumption. However, the deployment of edge systems often assumes the availability of low-latency communication infrastructure such as 5G or fiber networks. These assumptions do not hold universally, particularly in areas with limited coverage or with dense population \cite{horvath20256g}. Wi-Fi can provide an alternative in densely populated areas \cite{huerta2023long}. Furthermore, many urban monitoring (and smart city) needs are inherently local, targeting specific intersections, pedestrian zones, or parking facilities. In such cases, wide-area connectivity may be unnecessary or even counterproductive.

This work proposes a localized edge AI framework that uses Wi-Fi as a primary communication technology to connect \textbf{low-power IoT camera sensors} with shared edge devices to offload AI inference on vehicle classification. 

Each camera node captures high-resolution images and processes them locally using its CPU or offloads them to Nvidia Jetson Nano nodes for GPU-accelerated inference. The design prioritizes nearby, high-performance processing within the coverage range of a Wi-Fi network, eliminating reliance on wide-area infrastructure.

The main contributions of this work are: 
\begin{itemize}
    \item Design and implementation of a Wi-Fi–based edge AI architecture that connects camera sensor nodes to edge devices for real-time image classification.
    \item Evaluation of the feasibility of local versus offloaded AI inference for vehicle classification in smart cities.
    \item Quantification of trade-offs among processing performance, dropped frames, and inference latency in single-node deployments.
\end{itemize}

This paper is organized as follows: Section~\ref{sec:relatedWork} reviews related work on wireless technologies for smart cities, highlighting trade-offs between 5G and Wi-Fi in urban deployments. Section~\ref{sec:architecture} presents our system's architecture and experimental setup. Section~\ref{sec:EvaluationDesign} defines the performance metrics used in our evaluation. Section~\ref{sec:Results} reports and discusses the results of our experiments. 
Section~\ref{sec:Future} discusses outcomes and describes challenges still needing to be adressed.
Section~\ref{sec:Conclusion} concludes the paper.

\section{Related Work}\label{sec:relatedWork}

Understanding the general characteristics of the edge communication and compute infrastructure is essential for supporting smart city applications. Therefore, the work of Chen et al. \cite{9430904} contributes to understanding the role of Mobile Edge Computing (MEC) in enabling smarter cities. The authors highlight MEC’s ability to support latency-sensitive services by placing computation close to the data source, emphasizing leveraging 5G networks to achieve the required bandwidth and low latency. Their architecture envisions MEC nodes at the network edge, enabling applications such as real-time video analytics, augmented reality, and vehicular communication. However, the reliance on 5G infrastructure limits its immediate applicability in areas without widespread 5G coverage or where deployment costs are prohibitive.

Similarly, Garcia et al. \cite{garcia2018wireless} analyze wireless technology selection for IoT deployments in smart cities, focusing on how technology choice impacts reliability and performance. They identify overcrowding in the 2.4 GHz ISM band due to the widespread use of Bluetooth, ZigBee, Wi-Fi, mobile communications, beeing a challenge for Worldwide Interoperability for Microwave Access (WiMax), causing interference, packet loss, and degraded performance. Their findings suggest that technology selection must balance application-specific requirements with interference mitigation, making Wi-Fi a viable but context-dependent alternative to cellular-based solutions.

The work of Taleb et al. \cite{7876955} provides a comprehensive review of smart city architectures, identifying wireless technologies as key enablers of urban IoT services. They recognize both 5G and Wi-Fi as critical in urban deployments, noting that while 5G excels in wide-area, high-mobility scenarios, Wi-Fi offers cost-effective coverage for ultra-local services where infrastructure investment must remain minimal.

Together, these studies emphasize that although 5G promises high performance for large-scale and mobile smart city services, Wi-Fi remains a practical alternative in local or budget-constrained scenarios. Reflecting on the findings by Garcia \cite{garcia2018wireless} from 2018, shifting Wi-Fi into the 5/6 Ghz band could resolve many of the problems mentioned. This motivates our exploration of a Wi-Fi-based edge AI framework for real-time traffic monitoring, removing the dependency on wide-area infrastructure while meeting low-latency requirements.


\section{Framework}\label{sec:architecture}

In this section, we introduce the framework of a Smart City use case application for performing vehicle classification directly at the roadside, proposing a low-power, low-cost solution, and propagating the information over short range to client applications via Wi-Fi.

\begin{figure}[ht]
    \centering
    \vspace{-2mm}
    \includegraphics[width=0.45\textwidth]{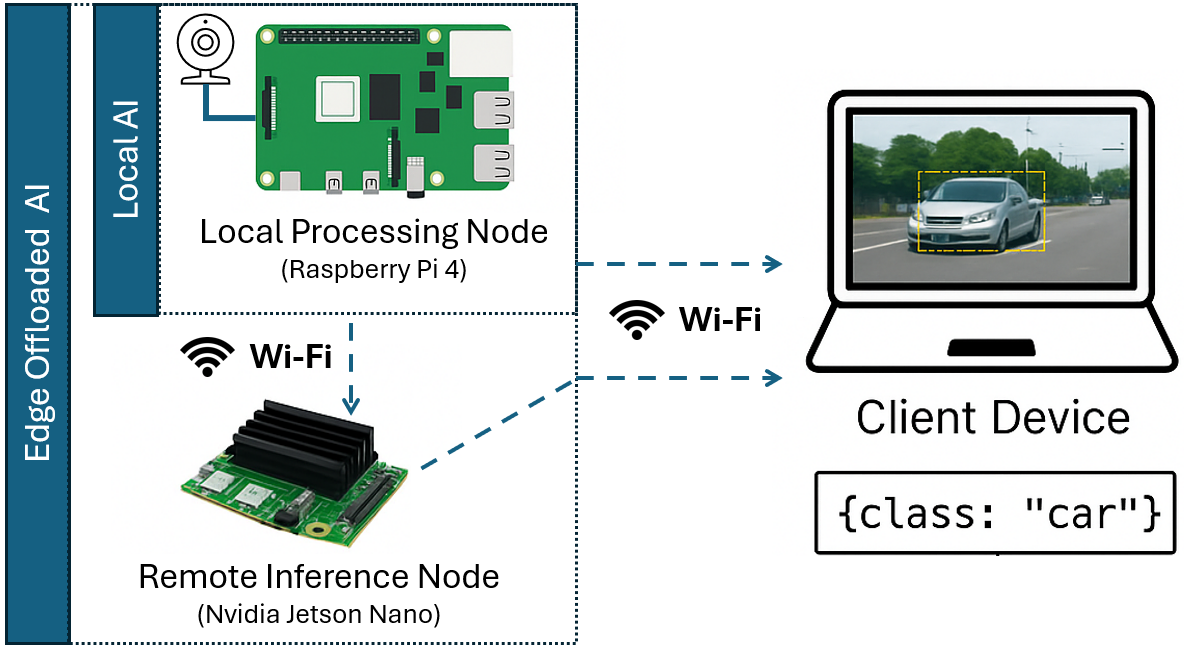} %
        \vspace{-3mm}
    \caption{Overview of the system architecture}%
    \Description{The core components include Local AI, Offloaded AI, and the client, all connected via Wi-Fi. The Local AI node, using a Raspberry Pi, handles image acquisition.}
    \label{fig:arch}%
    \vspace{-4mm}
\end{figure}

As depicted in Figure~\ref{fig:arch}, the architecture of the framework consists of three main components, client, local processing node, and remote inference node, deployed in two different configurations optimized for different objectives, namely \textbf{local} and \textbf{offloading}. The first deployment, referred to as the \textbf{local deployment}, involves a node performing vehicle classification directly on the roadside using a Raspberry Pi 4, which also acquires the input images. This setup uses CPU-only inference to achieve entirely local processing. A client device connects to the local node via Wi-Fi and requests image classification results at regular intervals. The results include annotated vehicle classifications in JSON format and image overlays (see Section~\ref{subsec:vehicle}), providing the most recent outcomes to the client applications.   

In the \textbf{offloading deployment}, inference is performed remotely on an edge node equipped with a GPU (NVIDIA Jetson Nano) for acceleration. When the client requests it, the local node sends the current image to the remote node, executing the image classification. After processing, the client device retrieves the results from the remote node. 

This setup directly compares local (CPU-based) and offloaded (GPU-accelerated) edge processing, quantifying the impact of CPU and GPU inference performance in conjunction with additional networking latency introduced by the offloading process.

\subsection{Workflow}\label{subsec:workflow}
This section describes how the three main components interact in the depicted deployment configurations.
Based on the deployment configuration, the local processing node conducts image acquisition and inference, where the remote inference node only conducts inference on images provided by the local processing node. The components organize their communication through the following messages:
\begin{itemize}
    \item \textbf{Hello(DeviceId)}: Announces device presence to the client
    \item \textbf{Control}: Comprises multiple subtypes:
    \begin{itemize}
        \item \textit{ConfigureExperiment}: Declares model, operational mode, and duration of the experiment
        \item \textit{ReadyToStart}: Confirms model loaded and warm-up completed
        \item \textit{BeginExperiment}: Initiates periodic pulsing
        \item \textit{Shutdown}: Performs orderly teardown
    \end{itemize}
    \item \textbf{Pulse(TimingMetadata)}: Emitted by client every $\Delta t$
    \item \textbf{Frame}: Conveys image $F_i$ from local to remote node
    \item \textbf{Result}: Returns inference output $R_i$ to client
\end{itemize}

Both \textbf{Frame} and \textbf{Result} messages carry the \texttt{TimingMetadata} package initially sent by the \textbf{Pulse}. This package contains the sequence number $i$ of the current inference cycle, the frame number, and a set of timestamps collected across devices. 

The use of individual messages is described in the following sections.

\subsubsection{Workflow local inference}
\begin{figure}[ht]
    \centering
            \vspace{-4mm}
    \includegraphics[width=0.30\textwidth]{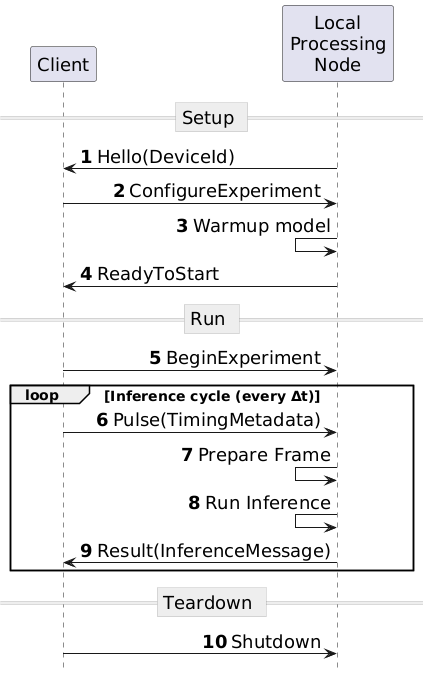}
         \vspace{-4mm}
    \caption{workflow on local deployment configuration using local inference}
         \Description{Workflow for local processing, showing interactions between client and local processing node.}
              \vspace{-4mm}
    \label{fig:wfLocal}

\end{figure}

 is depicted in Figure~\ref{fig:wfLocal} and initiated by the client in a \textit{Setup}-phase with the local processing node.
In the \textit{Setup}-phase, the client awaits \texttt{Hello} messages from the local processing node, then issues \texttt{Control::ConfigureExperiment} to configure the model, in the current deployment (local or offloading), and defines run parameters (e.g., experiment duration, fixed frame rate).
The local processing node loads the model, performs a warm-up cycle, loading the model into memory and executing a dummy inference pass to prepare the runtime, and then replies to the client with \texttt{Control::ReadyToStart}.

The \textit{Run}-phase starts with the \texttt{Control::BeginExperiment} message indicating a new run. 
For each inference cycle, the client emits a \texttt{Pulse} message carrying a \texttt{TimingMetadata} package that includes the current sequence number $\text{i}$.
Upon receiving a \texttt{Pulse}, the local processing node captures image $F_i$ from the onboard camera. The system packs the captured frame into a \texttt{Frame($F_i$)} message.Upon capturing, the node applies inference and returns the result to the client as \texttt{Result($R_i$)}.
The client completes the workflow in the \textit{Teardown} phase by submitting the \texttt{Shutdown} message.

\subsubsection{Workflow offloaded inference}
\begin{figure}[ht]
    \centering
    \includegraphics[width=0.44\textwidth]{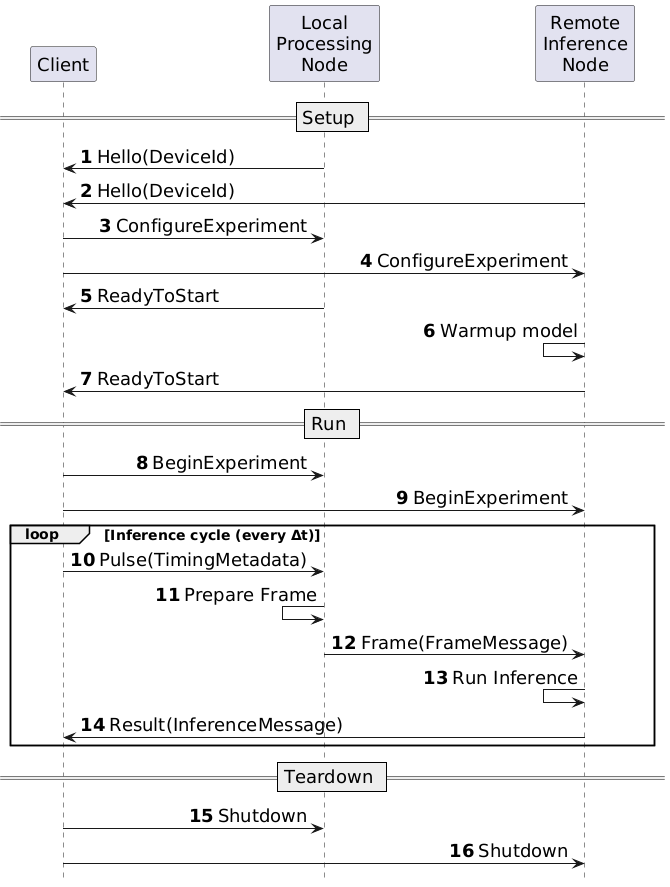}
    \vspace{-4mm}
    \caption{workflow on offloading deployment configuration applying remote inference}
     \Description{Workflow for offloaded processing, showing interactions between client, local, and remote processing nodes.}
        \vspace{-2mm}
    \label{fig:wfremote}

\end{figure}

is depicted in Figure~\ref{fig:wfremote}, and extends on the \textbf{local inference deployment}. During the \textit{Setup}-phase, the client also registers the remote inference node by awaiting \texttt{Hello} and then sending the \texttt{Control::ConfigureExperiment} message. Both nodes respond with \texttt{Control::ReadyToStart}.

During the \textit{Run}-phase, the local processing node transmits \\ \texttt{Frame($F_i$)} to the remote inference node.
The remote inference node executes the model to obtain the inference result $R_i$. The remote node then returns \texttt{Result($R_i$)} to the client.
This deployment leverages GPU acceleration at the cost of network transmission overhead, such that the total processing time includes inference and data transfer delays.

The run continues until the configured experiment duration has elapsed, after which the client initiates the \textit{Teardown}-phase by sending \texttt{Control::Shutdown} to both nodes, thus terminating the experiment.

\subsubsection{Distributed offloading}\label{subsec:distoffload}
The proposed architecture is not necessarily constrained to a bijective relationship between the local processing node $N^{loc}$ and remote inference node $N^{rem}$. Generalizing the queueing concept $N^{rem}$ to support $1:k$ node relations, where $k$ is the number of $N^{loc}=\{n_0 \dots n_{k-1}\}$ share one $N^{rem}$ node for offloading.

With the capacity queue policy ($\forall n \in N^{loc}: Q_{\text{size}}=1$), each local processing node supports only the most current frame to be processed. 

We must demand the following condition to assign $n_i \in N^{loc}$.
\[
    \frac{\mu_m}{\sum_{i=0}^k\lambda_0(n_i)} \;\ge\; 1,
\]
where $\mu_m$ is the highest sustainable processing rate for model $m$ and $\lambda_0$ defines per-node frame generation rate.

\subsection{Vehicle Classification using AI}\label{subsec:vehicle}
In this section, we describe the purpose of the framework.
Many traffic applications widely use these existing AI models for live traffic classification \cite{rodriguez2022analysis} \cite{zhang2022real} \cite{maity2021faster}. We also utilize \emph{YOLOv5} \cite{ultralytics_yolov5} as object-detector. The models are trained on the COCO dataset \cite{luo2018mio, zuraimi2021vehicle} and on car, truck, bus, and motorcycle categories.

Three model distinguished by their size: \texttt{YOLOv5n} (nano), \texttt{YOLOv5s} (small), and \texttt{YOLOv5m} (medium). 
\texttt{YOLOv5n} variant provides the lowest latency but reduced precision, while \texttt{YOLOv5m} achieves higher accuracy at greater computational cost.

This selection enables systematic evaluation on both the \emph{Local Processing Node} (CPU-only) and the \emph{Remote Inference Node} (GPU-accelerated), showing a trade-off between processing locality. In terms of model size, their parameter counts and corresponding in-memory weights differ substantially, about \emph{7 MB} for the nano-size model, \emph{29 MB} for the medium, and \emph{85 MB} at FP32 precision. The official YOLOv5 model table \cite{ultralytics_yolov5} reports the parameter counts that we use to derive these values, applying the relation of $\text{params} \times 4 \,\text{bytes}$ for FP32 storage.

\begin{figure}[!ht]
\centering
\begin{subfigure}{0.47\linewidth}
    \includegraphics[width=\linewidth]{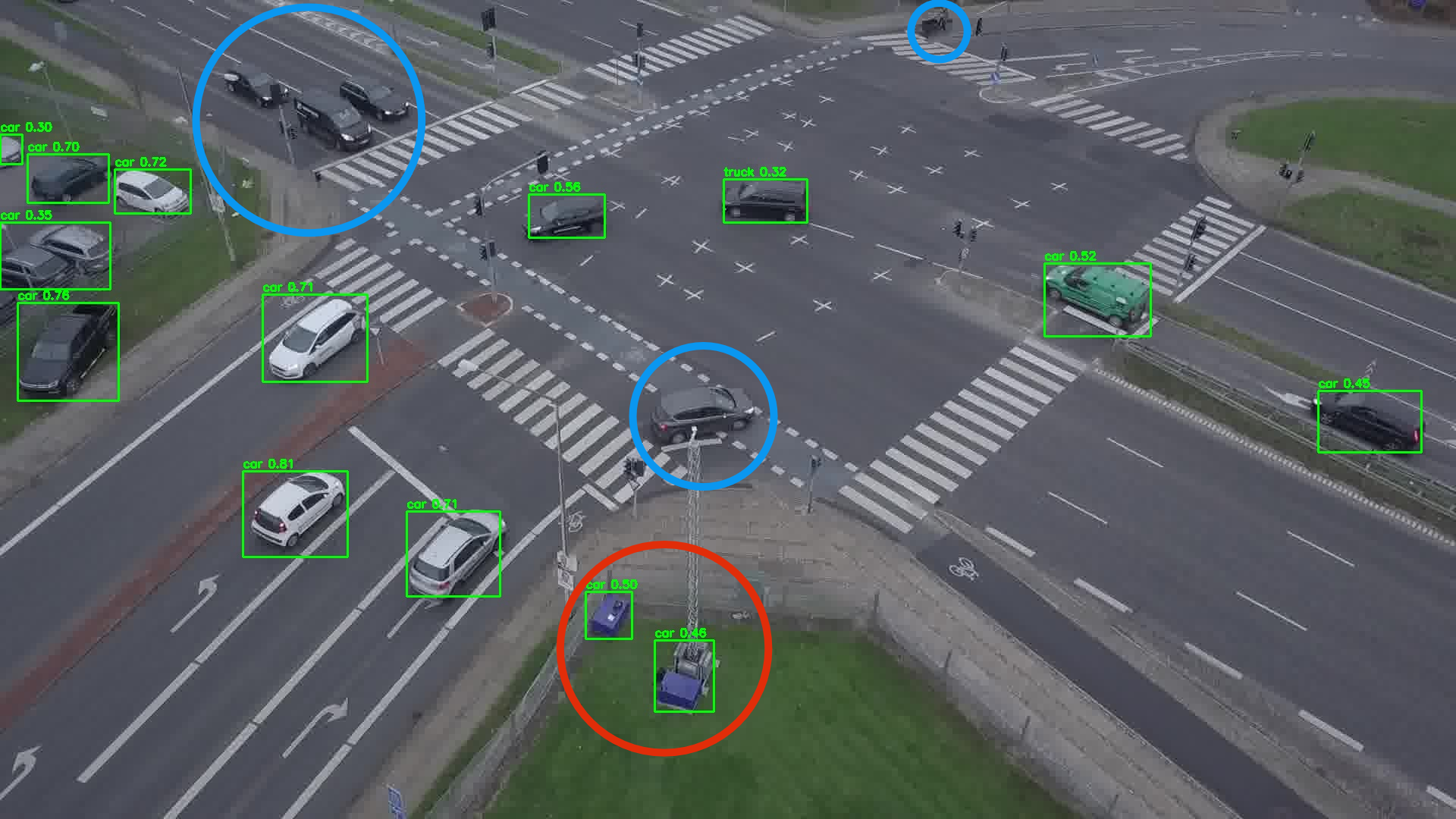}
    \caption{Frame 30175; YOLOV5n}
    \label{fig:yolo-n-1}
\end{subfigure}\hfill
\begin{subfigure}{0.47\linewidth}
    \includegraphics[width=\linewidth]{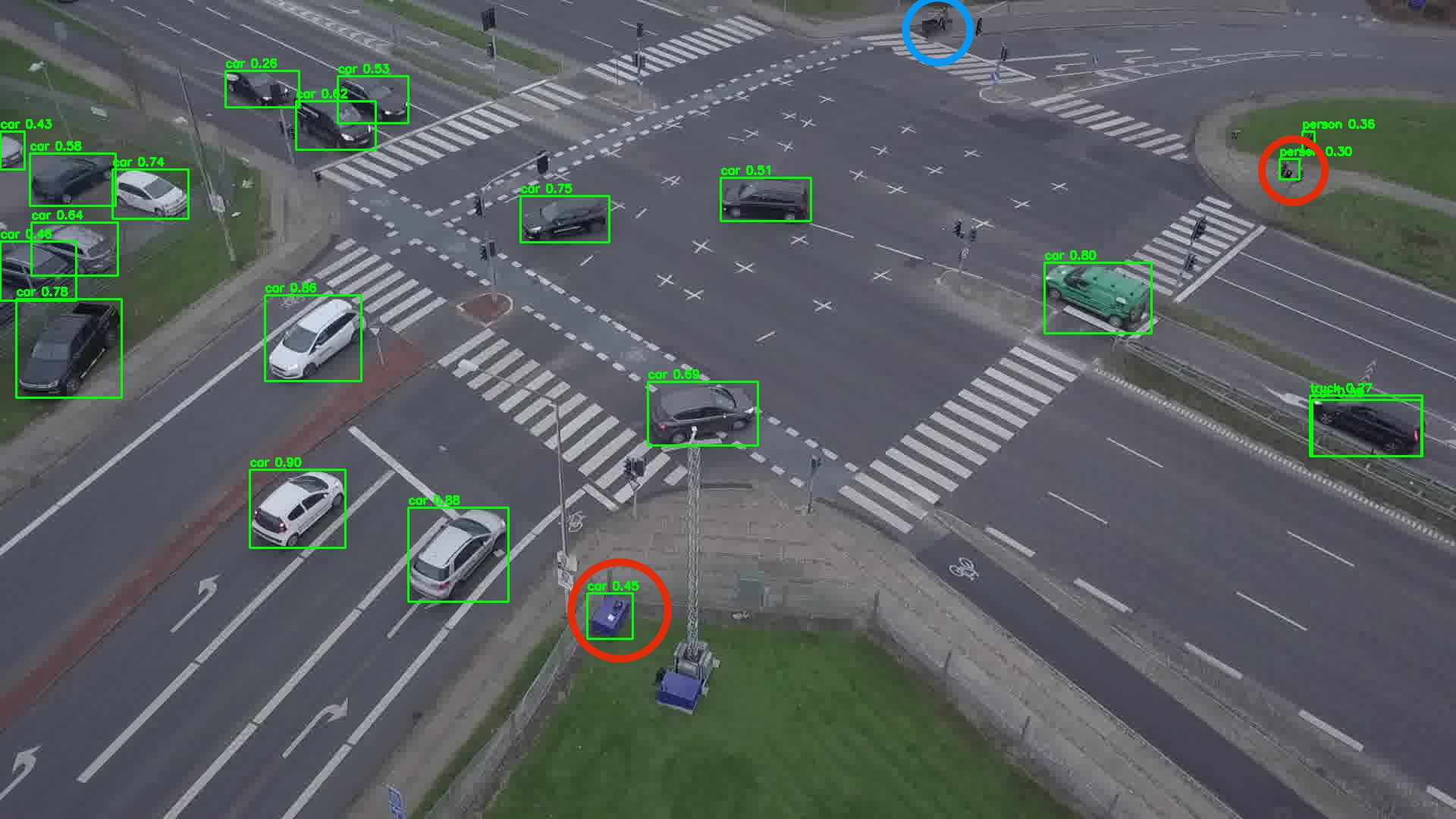}
    \caption{Frame 30175; YOLOV5m}
    \label{fig:yolo-m-1}
\end{subfigure}

\begin{subfigure}{0.47\linewidth}
    \includegraphics[width=\linewidth]{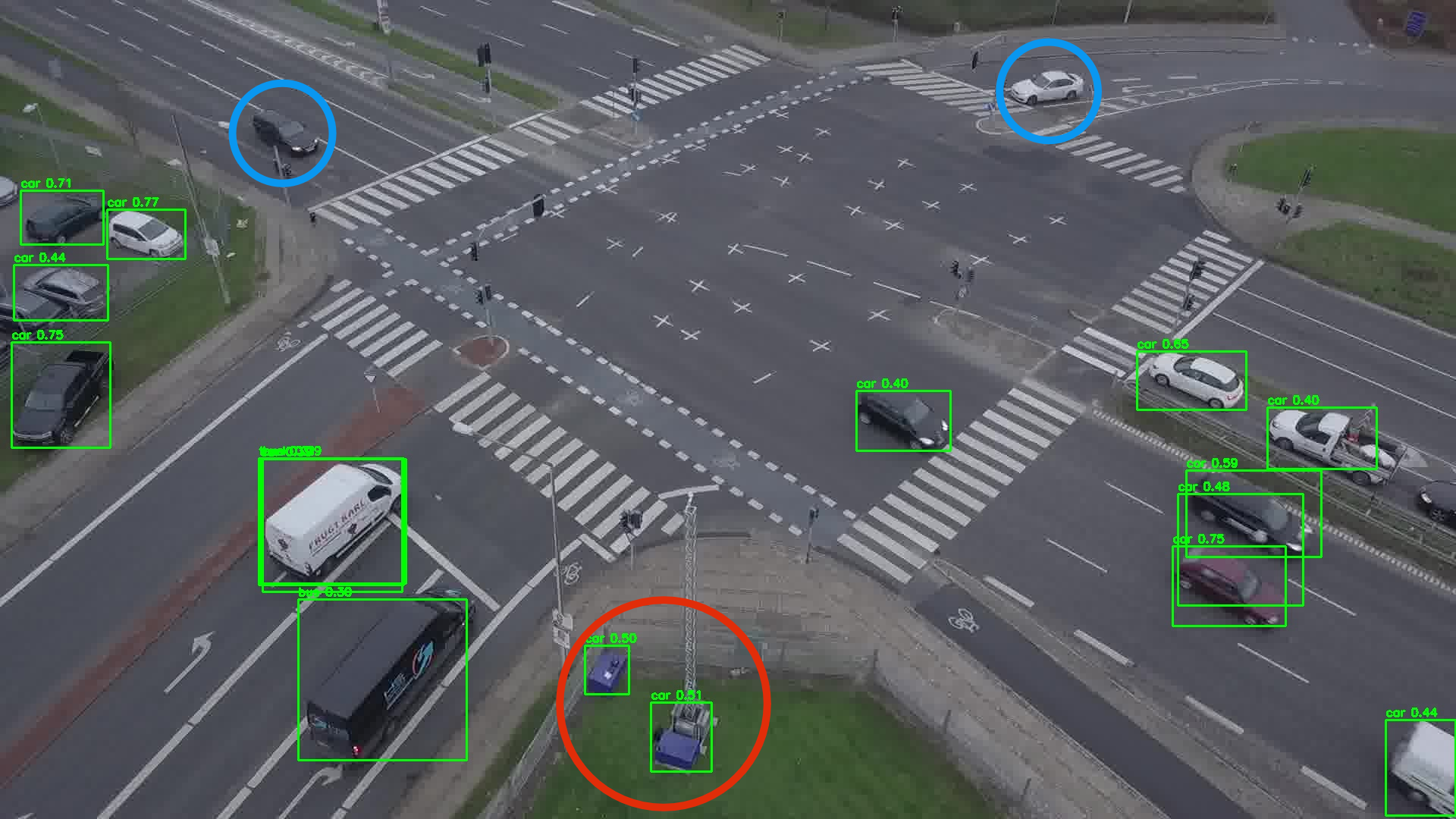}
    \caption{Frame 24418; YOLOV5n}
    \label{fig:yolo-n-2}
\end{subfigure}\hfill
\begin{subfigure}{0.47\linewidth}
    \includegraphics[width=\linewidth]{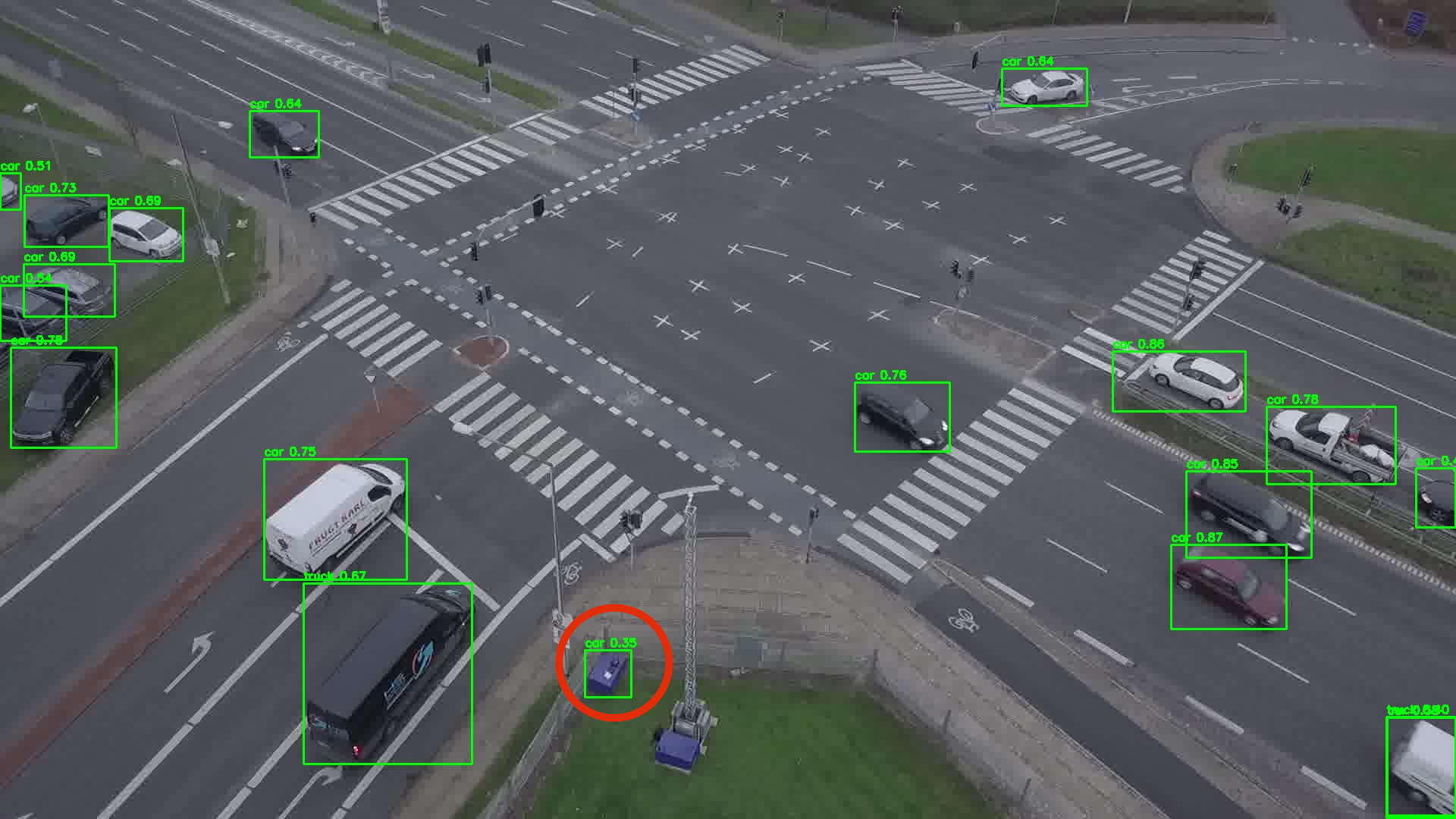}
    \caption{Frame 24418; YOLOV5m}
    \label{fig:yolo-m-2}
\end{subfigure}

\begin{subfigure}{0.47\linewidth}
    \includegraphics[width=\linewidth]{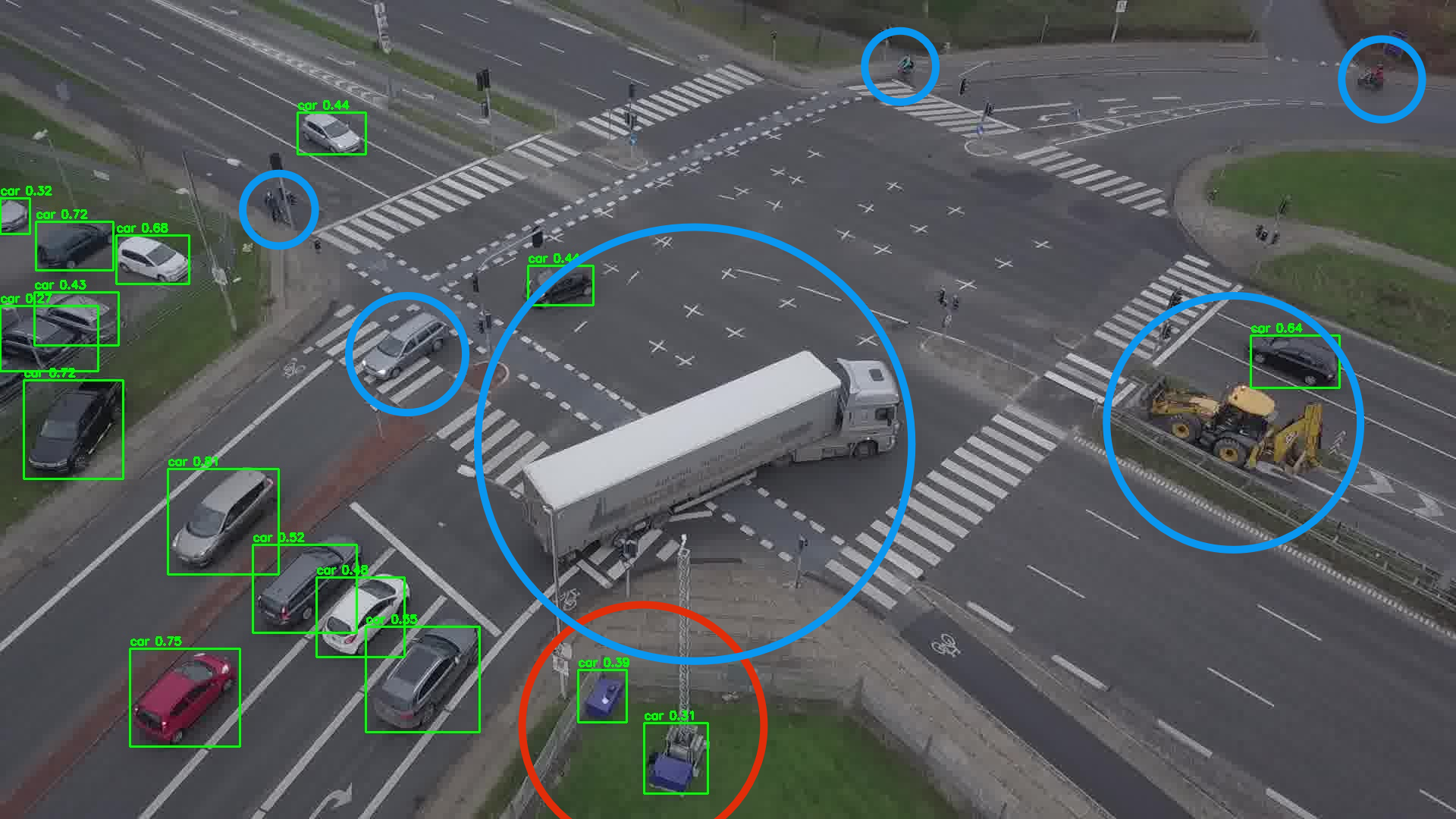}
    \caption{Frame 10614; YOLOV5n}
    \label{fig:yolo-n-3}
\end{subfigure}\hfill
\begin{subfigure}{0.47\linewidth}
    \includegraphics[width=\linewidth]{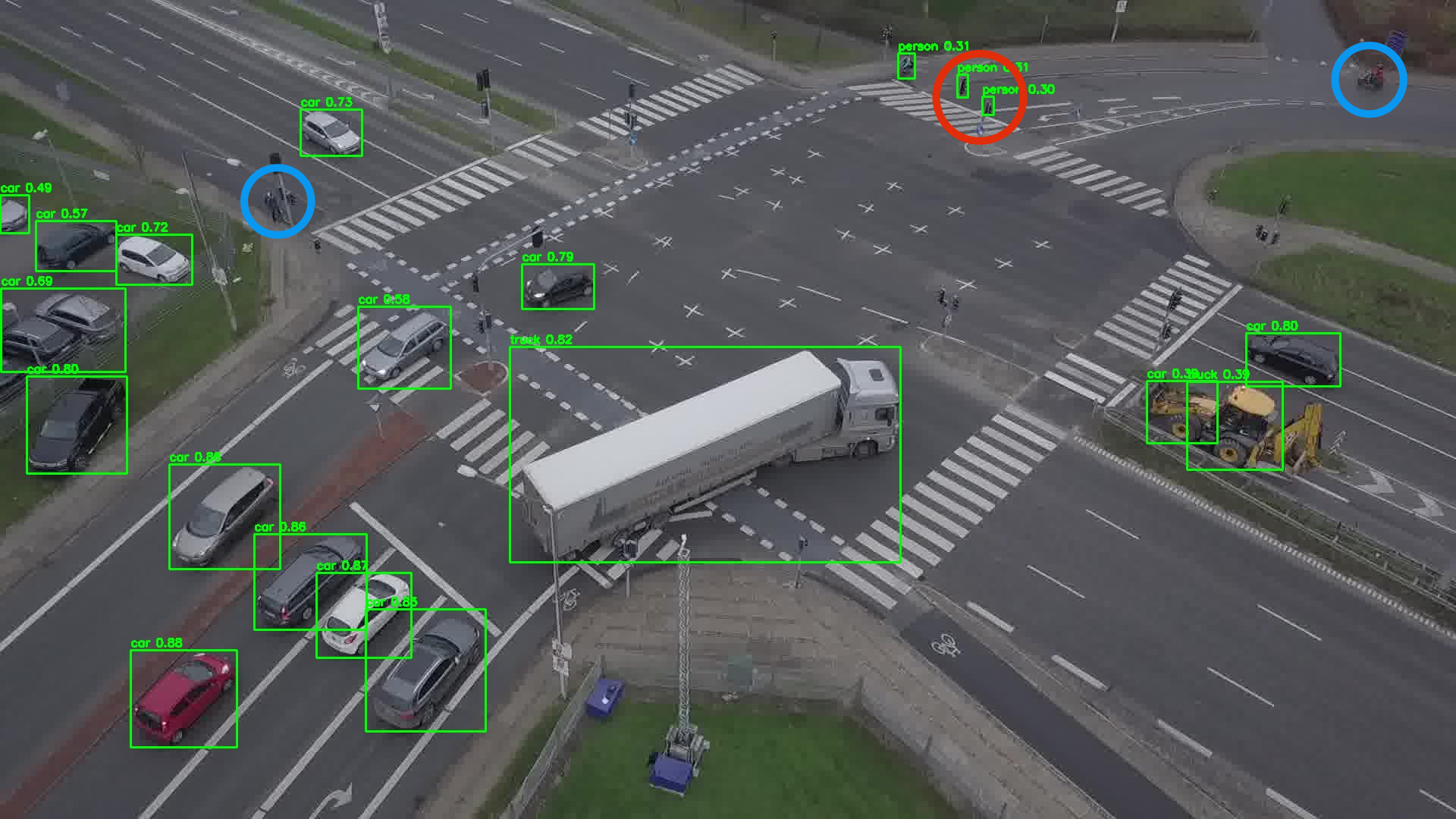}
    \caption{Frame 10614; YOLOV5m}
    \label{fig:yolo-m-3}
\end{subfigure}
\caption{comparison of \texttt{YOLOv5n} and \texttt{YOLOv5m} on frames from the multi-view traffic intersection dataset~\cite{mogelmose_multiview_traffic2019} 
}
\Description{Sample images comparing YOLOv5n and YOLOv5m, with red highlights indicating misclassifications and blue highlights indicating missed detections.}
        \vspace{-6mm}
\label{fig:yolo-overview}
\end{figure}

Figure~\ref{fig:yolo-overview} compares the outputs of \texttt{YOLOv5n} and \texttt{YOLOv5m} on identical frames. 
Both models detect multiple vehicles, but they also misclassify some objects. For instance, they identify roadside electrical boxes as cars, and in the case of \texttt{YOLOv5m}, they also misclassify traffic lights as pedestrians, as highlighted in red in Figures \ref{fig:yolo-n-1}, \ref{fig:yolo-n-2}, and \ref{fig:yolo-n-3}.
\texttt{YOLOv5n} further fails to detect larger objects such as the truck and excavator (highlighted in blue; Figure ~\ref{fig:yolo-n-3}), whereas the medium variant succeeds in recognizing them and generally identifies more vehicles overall. 
These examples illustrate the trade-off between accuracy and model size \cite{menghani2023efficient}. Smaller models are computationally efficient but prone to missed detections (Figure ~\ref{fig:yolo-n-1}), while larger models achieve broader coverage at the expense of higher computational cost (Figure ~\ref{fig:yolo-m-3}).

\subsection{Implementation}\label{subsec:impl}
The solution implements a distributed edge computing architecture with four main software components implemented in Rust and Python. The code applies the actual inference in Python. The complete implementation is available on GitHub on our rust-traffic-watch repository\footnote{\url{https://github.com/rawalcher/rust-traffic-watch}}

The implementation supports automated benchmarking across multiple AI model variants (YOLOv5n/s/m), configurable frame rates (1-15 frames per second), and supports local- and offloading deployment configurations established in the following software components:
\begin{itemize}
    \item \textbf{Controller}: Orchestrates experiments, coordinates device communication, and collects performance metrics across all distributed nodes
    \item \textbf{Pi Sender}: Captures frames and performs local PyTorch inference or forwards data for offloading to remote compute nodes
    \item \textbf{Jetson Receiver}: Provides accelerated TensorRT inference for offloaded workloads with optimized GPU utilization
    \item \textbf{Shared Library}: Common networking protocols, binary serialization, and type definitions for inter-device communication
\end{itemize}


All experiments generate detailed CSV logs containing end-to-end latency, processing overhead, network latency, and detection accuracy metrics for reproducible analysis across 24 configurations.

\subsubsection{Hierarchical Time Measurement}
Accurate time measurements across distributed computing nodes always present difficulties since clocks tend not to be fully synchronized. This paragraph describes the method we use to avoid this problem.

We measure time hierarchically across nodes $\mathcal{N} = \{N_c, N_p, N_r\}$, where $N_c$ is the client, $N_p$ a local processing node, and $N_r$ a remote inference node triggered by $N_p$. Each node $N_i$ maintains a local clock. The client $N_c$ initializes the ordered timestamp structure $\mathcal{T}$ (\texttt{TimingMetadata}):
\[
\mathcal{T} = \langle t_1, \dots, t_n \rangle,
\]
During execution, $N_p$ appends timestamps for local processing and triggers $N_r$, which appends inference timestamps. We compare only timestamps acquired from the same local clock to avoid aligning the local clocks ($C_c, C_p, C_r$). So we define a duration between to moments $\tau_a,\tau_b$ on node $N_i$ as
\[
\Delta T_i(t_a, t_b) = C_i(\tau_b) - C_i(\tau_a),
\]
with $t_a = C_i(\tau_a)$ and $t_b = C_i(\tau_b)$. A valid duration requires:
\[
\Delta T_i(t_a, t_b) \quad \text{only if} \quad t_a, t_b \leftarrow C_i,
\]
i.e. to calculate the duration of a run is calculated as follows: 
\begin{align*} 
t_{\text{start}} = C_c(\tau_{\text{start}}),\\
t_{\text{end}} = C_c(\tau_{\text{end}}),\\
\Delta T_{\text{run}} = t_{\text{end}} - t_{\text{start}}.
\end{align*}


\section{Evaluation Design}\label{sec:EvaluationDesign}

In our evaluation, we systematically vary the capture frame rate $CFR$ from 1 to 15 frames per second (FPS) and measure the impact of different inference models on total processing time, accuracy, and system responsiveness. These parameters allow us to analyze trade-offs between computation location, network usage, and classification performance, providing insights into the design of smart city edge AI systems, considering the two different implementations of the vehicle classification smart city application.

\subsection{Evaluation Metrics}\label{subsec:metrics}
Table \ref{tab:notation} summarizes the notation used throughout the evaluation. This table serves as a reference for the various symbols and terms employed in the evaluation.

\begin{table}[h]
\vspace{-3mm}
\caption{Notation used in describing the experiments}
\vspace{-4mm}
\label{tab:notation}
\centering
\scalebox{0.9}{
\begin{tabular}{lll}
\toprule
\textbf{Symbol} & \textbf{Description} & \textbf{Unit} \\
\midrule
$CFR$ & Capture frame rate & \si{frames\per\second} \\
$DFR$ & Dropped frame rate & \si{frames\per\second} \\
$OTR$ & On-time processing rate & \si{frames\per\second} \\
$RPI$ & Relative performance improvement & - \\
$\mathcal{Z}_{\text{total}}$ & Total nominal frames captured during experiment & frames \\
$\mathcal{Z}_{\text{dropped}}$ & Amount of frames dropped during experiment & frames \\
$\Delta T_{\text{total}}$ & Total experiment duration & s \\
$\Delta T_{\text{proc}}(i)$ & Processing time for frame $i$ & s \\
$\mu_m$ & Max.\ sustainable processing rate for model $m$ & \si{frames\per\second} \\
$\lambda_0$ & Per-node frame generation rate & \si{frames\per\second} \\
$Q_{\text{size}}$ & Capacity queue size & frames \\
\bottomrule
\vspace{-8mm}
\end{tabular}

}
\end{table}

\subsubsection{Performance Metrics}
The target metric in our evaluation to assess image inference is the \emph{capture frame rate} ($CFR$), defined as:
\[
    CFR = \frac{\mathcal{Z}_{\text{total}}}{\Delta T_\text{total}}.
\]

To handle varying processing speeds, we implement a \emph{capacity-bounded queue} with size $Q_{\text{size}} = 1$, ensuring only the most recent frame is processed.
When a new frame arrives, it replaces any waiting frames in the queue during ongoing processing. This policy prioritizes the timeline over completeness.

We define the \emph{dropped frame rate} ($DFR$) as the ratio of frames discarded due to processing delays:
\[
    DFR = \frac{\mathcal{Z}_{\text{dropped}}}{\mathcal{Z}_{\text{total}}} \times 100\%.
\]

A frame is considered dropped when:
\begin{enumerate}
    \item It arrives while the previous frame is still being processed $\Delta T_{proc}(i-1) > 1/CFR$, and
    \item It gets replaced by a newer frame before processing begins.
\end{enumerate}

The \emph{on-time processing rate} ($OTR$) represents successfully processed frames, $OTR = 100\% - DFR$.

\subsubsection{Processing Time}

The \emph{average processing time per frame} $T_{proc}$ is computed as:
\vspace{-4mm}
\[
    \Delta T_{proc} = \frac{1}{n} \sum_{i} \Delta T_{proc}(i).
\]
Where $n$ represents the number of frames that the system successfully processes without dropping, while $i$ indicates the index of the current frame, which determines the processing time $\Delta T_{proc}$ for that frame, how processing time is cummulated depends on the deployment configuration. $\Delta T_{proc} \in \{\Delta T_{proc}^\text{loc}, \Delta T_{proc}^\text{off}\}$ 

For the \textbf{local-deployment}, processing time $\Delta T_{proc}^{loc}$ accumulates
Image acquisition from the camera, model inference, and result serialization.

For the \textbf{offloading-deployment} the processing time $\Delta T_{proc}^\text{off}$ includes additional components: network transmission $t_{in}$ to the remote node, the queuing delay at the remote node, model inference $\Delta T_{inf}$ and network transmission of results $\Delta T_{out}$.

\subsubsection{Performance Comparison}
To assess the benefit of GPU offloading, we define the \emph{relative performance improvement} ($RPI$):
\[
    RPI = \frac{DFR_{\text{local}} - DFR_{\text{offload}}}{DFR_{\text{local}}} \times 100\%.
\]

If $RPI > 0$, offloading reduces frame drops, \\
$RPI < 0$ shows that network overhead outweighs acceleration benefits. $RPI = 0$, both approaches work identically.

\subsubsection{Model-Specific Processing Capacity}
Each model variant $m \in \{\texttt{YOLOv5n}, \texttt{YOLOv5s}, \texttt{YOLOv5m}\}$
exhibits a maximum sustainable processing rate $\mu_m$ for each type $t$ processing node and corresponding hardware capabilities.
The processing capacity without drops demands:
\vspace{-4mm}
\[
    CFR \leq \mu_m.
\]

Due to the capacity queue implementation, the system can operate at higher capture rates but will drop frames according to:
\[
    DFR(m) \approx \max\left(0, (1 - \frac{\mu_m}{CFR})\right) \times 100\%.
\]

This approximation assumes steady-state operation and uniform processing times.

\section{Results}\label{sec:Results}
We evaluate two deployment strategies: \emph{local AI} inference directly on the roadside node (Raspberry~Pi~4, CPU-only) and \emph{remote inference} offloaded to a Jetson~Nano (GPU-enabled) over a Wi-Fi link. Performance is quantified using the metrics introduced in Section~\ref{sec:EvaluationDesign}: the \emph{Dropped Frame Rate} (\(DFR\)), the \emph{Capture Frame Rate} (\(CFR\)), and the \emph{Inference Latency} (\(\Delta T_{\text{inf}}\)).

 \begin{figure}[ht]
     \centering
     \vspace{-0mm}
     \includegraphics[width=0.45\textwidth]{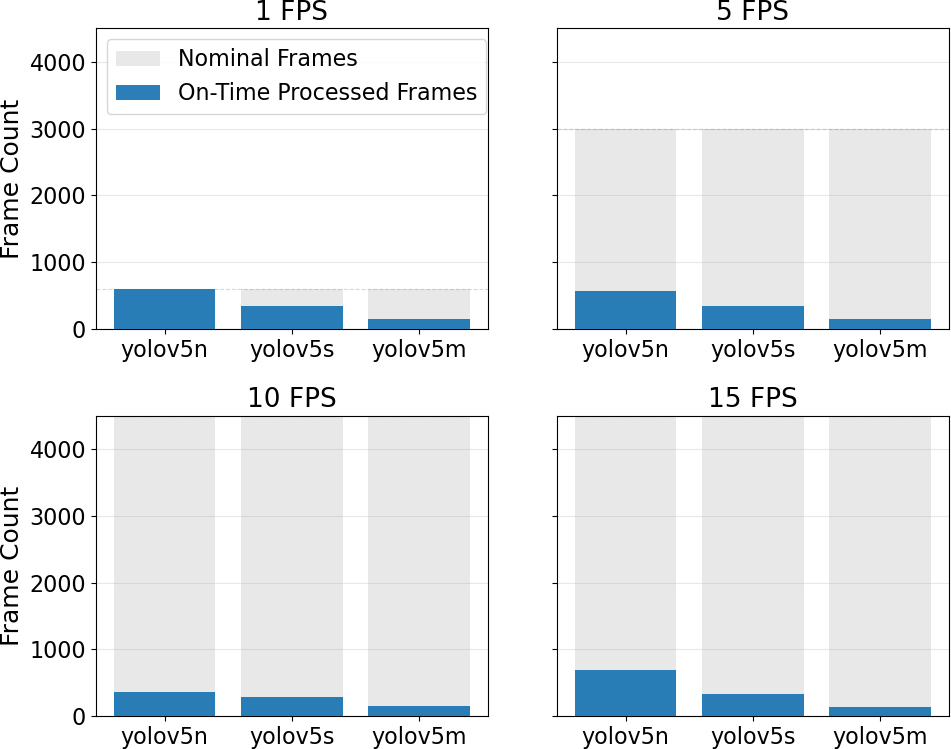} %
     \vspace{-2mm}
     \caption{OTR over DFR using local deployment configuration (y-axis limited to 4000 frames)}%
    \Description{Bar-chart comparing different frame rates using local deployment}
     \label{fig:local}%
         \vspace{-2mm}
 \end{figure}



Reviewing the results of local deployment configuration shows that sustained operation without frame drops (\(DFR = 0\)) is only achievable with the smallest model, YOLOv5n, at \(CFR = 1\). This setting satisfies the condition \(OTR \geq CPR\), so the next frame arrives only after the previous one is fully processed. Increasing the capture rate to \(5\)~FPS or \(15\)~FPS results in a significant reduction of successful processings, with \(DFR\) exceeding \(80\%\) for YOLOv5s and \(95\%\) for YOLOv5m. In these cases, the high \(DFR\) reflects the condition \(OTR \ll CFR \), where most frames cannot be processed in time.

  \begin{figure}[ht]
     \centering
     \vspace{-2mm}
     \includegraphics[width=0.45\textwidth]{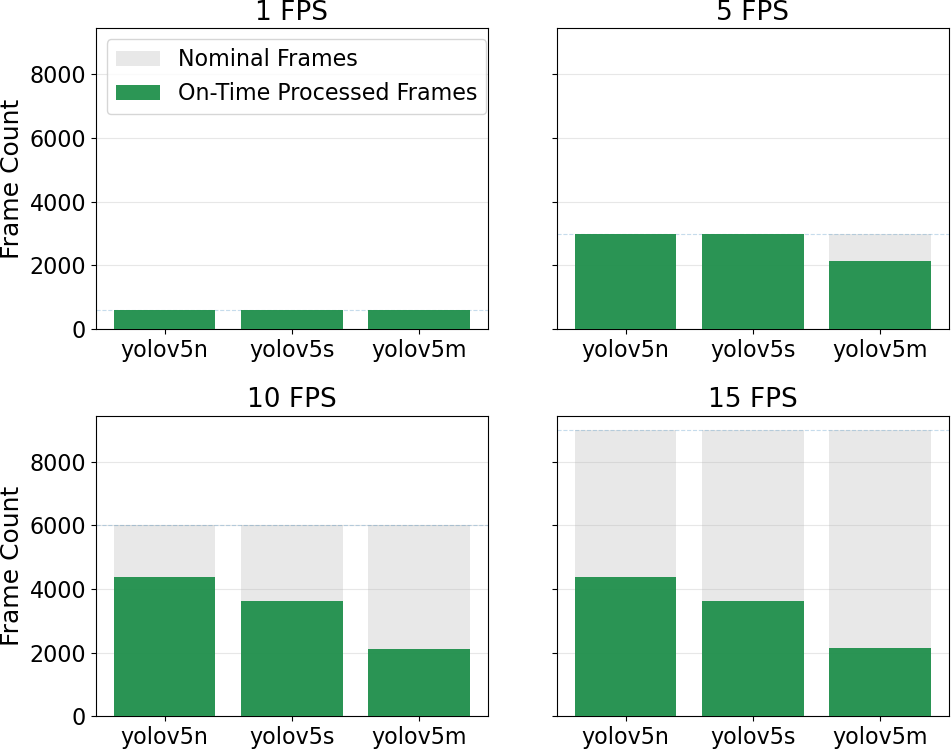} %
     \vspace{-2mm}
     \caption{OTR over DFR using offloading deployment configuration}%
      \Description{Bar-chart comparing for different frame rates using offload AI}
     \label{fig:offloaded}%
         \vspace{-2mm}
 \end{figure}

On the other hand, the situation changes quite a bit when offloading. With $CFR = 1$, we stay at $DFR \approx 0$, and we only start dropping frames at $CFR = 5$ with the bigger YOLOv5m model. Starting at $CFR = 10$, all models perform worse across the board with $DFR$ ranging from $\approx 27\%$ (YOLOv5n) up to $\approx 64\%$ (YOLOv5m). At $CFR = 15$, we can see the same results as observed at $CFR = 10$, facing hardware processing limits on the remote inference node.


These findings confirm that for single-node deployments, local inference is feasible only for lightweight models and low capture rates. In contrast, offloading to GPU-equipped edge nodes enables near-real-time performance even for heavier models at higher capture rates, as long as \(OTR\) remains sufficiently close to \(CFR\) to keep \(DFR\) within the acceptable bound \(DFR^{max}\). Furthermore, on lower $CFR \leq 5 $ resource capabilities of the offloading node suffice to process data of multiple nodes.

\section{Discussion}\label{sec:Future}
Based on the experimental evaluation, we found that processing the video stream on a single node can be highly restrictive. Many real-world deployments of edge AI involve multiple processing nodes sharing one offloading node. Such a configuration introduces several open challenges, like resource contention, adaptive scheduling, and reliable network performance despite the actual multi-node distribution of service capabilities.

\emph{Multi-node scalability} allows us to optimize resource usage. Increasing the number of camera nodes will increase the total data rate and computational demand on the AI node \cite{hakani2024edge}, potentially overloading its GPU or saturating the Wi-Fi network (see Section \ref{subsec:distoffload}). Understanding the limits of performance degradation and identifying optimal load-balancing and queuing strategies will be critical. This also demands long-term reliability testing in outdoor urban environments with variable interference, integrating more advanced AI models for real-time object detection and tracking, and applying compression techniques to reduce bandwidth requirements without compromising classification accuracy.



Finally, \emph{network optimization} techniques will be necessary to mitigate interference and congestion in dense Wi-Fi environments. This includes channel allocation strategies, use of Wi-Fi 7 \cite{deng2020ieee} features like Multi-Link Operation (MLO) \cite{carrascosa2024performance} and 4K Quadrature Amplitude Modulation (4K-QAM) \cite{hoefel2024effects}, and integration with mesh networking to extend coverage without sacrificing performance or going beyond those aspects of 6G.

\vspace{-4mm}
\section{Conclusion}\label{sec:Conclusion}
This paper introduced a Wi-Fi-based Edge AI framework for real-time traffic image classification in smart city environments. We implemented a system where low-power camera nodes either perform on-device inference using their CPU or offload the processing to an NVIDIA Jetson Nano for GPU-accelerated inference. By defining formal performance metrics such as inference latency and dropped frame rate, we could assess the feasibility of both local and offloaded processing strategies.

Our results demonstrate that purely local processing is only practical for lightweight AI models at low frame rates. For instance, YOLOv5n at 1 FPS processed all frames without drops. However, increasing the frame rate or the model complexity quickly caused dropped frame rates to exceed operational thresholds up to 98\% in some configurations. Conversely, GPU offloading to the Jetson Nano maintained adequate processing at low to moderate frame rates across all tested models, with only minimal performance degradation for heavier models at higher frame rates.

These findings underscore the value of GPU-accelerated offloading over Wi-Fi for achieving reliable, low-latency performance in urban traffic monitoring scenarios \cite{wolniak2024artificial} \cite{luusua2023urban} without the need for costly 5G or wired infrastructure \cite{horvath2025enhancing}. The approach is particularly attractive for \textbf{ultra-local deployments} where coverage is contained within a single Wi-Fi cell, offering a cost-effective yet high-performance alternative to more infrastructure-heavy solutions. Beyond the specific traffic monitoring use case, the proposed framework demonstrates the broader viability of Wi-Fi-based edge computing for other urban sensing applications, provided the computational load and network conditions are appropriately managed. This positions localized Wi-Fi edge AI as a practical enabler for smart city services in settings where traditional wide-area edge or cloud deployments are impractical.

\section*{Acknowledgment}
\textbf{This work is a preprint! }
Official DOI: \href{}{\url{https://doi.org/10.34749/3061-1008.2025.7}}

This work received funding from the OeAD agency of the Austrian Ministry for Education, Science and Research and the Macedonian Ministry for Education and Science under project number MK10/2024.


\bibliography{BibFile}
\bibliographystyle{ACM-Reference-Format}
\end{document}